\documentclass[nofootinbib]{revtex4}
\usepackage{graphicx}
\usepackage{latexsym}
\def\be{\begin{equation}}
\def\ee{\end{equation}}
\def\bea{\begin{eqnarray}}
\def\eea{\end{eqnarray}}

\begin{document}
\title{Gravitational Leptogenesis and Neutrino Mass Limit}
\author{Hong Li}
\author{Mingzhe Li}
\author{Xinmin Zhang}
\affiliation{Institute of High Energy Physics, Chinese
Academy of Sciences, P.O. Box 918-4, Beijing 100039, People's Republic of China}

\begin{abstract}

Recently Davoudiasl {\it et al} \cite{steinhardt} have introduced
a new type of interaction between the Ricci scalar $R$ and the
baryon current $J^{\mu}$, ${\partial_\mu R} J^{\mu}$ and proposed
a mechanism for baryogenesis, the gravitational baryogenesis.
Generally, however, $\partial_{\mu} R$ vanishes in the radiation
dominated era. In this paper we consider a generalized form of
their interaction, $\partial_{\mu}f(R)J^{\mu}$ and study again the
possibility of gravitational baryo(lepto)genesis. Taking $f(R)\sim
\ln R$, we will show that $\partial_{\mu}f(R)\sim \partial_{\mu}
R/R$ does not vanish and the required baryon number asymmetry can
be {\it naturally} generated in the early universe.

\end{abstract}

\maketitle

\hskip 1.6cm
PACS number(s): 98.80.Cq, 11.30.Er
\vskip 0.4cm

The origin of the baryon number asymmetry remains a big puzzle in
cosmology and particle physics. Conventionally, it is argued that
this asymmetry is generated from an initial baryon symmetric phase
dynamically as long as the following conditions are satisfied
\cite{sak}: (1) baryon number non-conserving interactions; (2) $C$
and $CP$ violations; (3) out of thermal equilibrium. When the
$CPT$ is violated dynamically, however the baryon number asymmetry
can be generated in thermal equilibrium \cite{ck}. In connecting
to dark energy recently we have studied a class of models of
spontaneous baryo(lepto)genesis \cite{li,li2} \footnote{For
related studies, see \cite{trodden}} by introducing a interaction
between the dynamical dark energy scalars and the ordinary matter.
Specifically, we have considered a derivative coupling of the
quintessential scalar field $Q$ to the baryon or lepton current,
\begin{equation}\label{lagr}
{\cal L}_{int} \sim {\partial_{\mu} Q}J^{\mu}~.
\end{equation}
One silent feature of this scenario for baryogenesis is that the
present accelerating expansion and the generation of the matter
and antimatter asymmetry of our universe is described in a unified
way.

Recently, Davoudiasl {\it et al} \cite{steinhardt} proposed a new
mechanism (dubbed gravitational baryogenesis) of generating baryon
asymmetry in thermal equilibrium. They introduced explicitly an
interaction between the Ricci scalar curvature with derivative and
the baryon number current: \be
 \mathcal L= \frac{1}{M^2} \partial_{\mu}R J^{\mu}~.
 \ee
And the baryon number asymmetry is given by
\be
 \frac{n_B}{s}\sim \frac{\dot R}{M^2 T}~,
 \ee
which shows that $n_B/s$ is determined by the value of $\dot R$,
however the Einstein equation, $R=8\pi G T^{\mu}_{\mu}=8\pi G
(1-3w)\rho$, tells us that $\dot R=0$ in the radiation-dominated
epoch of the standard Friedmann-Robertson-Walker (FRW) cosmology.
Davoudiasl {\it et al} in Ref. \cite{steinhardt} have considered
three different possibilities of obtaining a non-vanishing $\dot
R$ which include the effects of trace anomaly, reheating and
introducing a non-thermal component with $w>1/3$ dominant in the
early universe. In the braneworld scenario Shiromizu and Koyama in
Ref. \cite{shiromizu} provided another example for $\dot R\neq 0$.

In this paper we propose a generalized form of the derivative
coupling of the Ricci scalar to the ordinary matter:
 \be\label{genint}
 {\cal L}_{int} \sim \partial_{\mu}f(R) J^{\mu}~,
 \ee
then study the possibility of gravitational baryo(lepto)genesis.
In Eq. (\ref{genint}), $f(R)$ is a function of $R$ and for a
detailed study in this paper we take explicitly $f(R)\sim \ln R$.
So we have now an effective Lagrangian
 \be\label{intlagr}
 {\cal L}_{int}=-c\frac{\partial_{\mu}R}{R} J^{\mu}~.
 \ee
This type of operators is expected by integrating out the heavy
particles or extra dimension \cite{turtro,randall} and the
coefficient $c$ characterizes the strength of this new interaction
in the effective theory. In Refs. \cite{turtro,randall} effective
operators like $1/(R^m)$ (with $m>0$) have been considered for the
purpose of modifying the Einstein gravity and dynamically solving
the cosmological constant problem. In this paper we will show that
within the framework of the standard FRW cosmology, our model with
$\mathcal{L}_{int}$ in (\ref{intlagr}) can naturally generate the
baryon number asymmetry in the early universe.

Taking $J^{\mu}=J^{\mu}_B$, during the evolution of the spatial
flat FRW universe, ${\cal L}_{int}$ in Eq. (\ref{intlagr}) gives
rise to an effective chemical potential $\mu_{b}$ for baryons:
\bea\label{mub} -\frac{c}{R}{\partial_{\mu}R} ~J^{\mu}\rightarrow
-c\frac{\dot R}{R}n_{B}=-c\frac{\dot R}{R}(n_{b}-n_{\bar{b}})~,\nonumber\\
\mu_{b}=-c\frac{\dot R}{R}=-\mu_{\bar{b}}~. \eea
In thermal
equilibrium, the net baryon number density doesn't vanish as long as
$\mu_b\neq 0$ (when $T\gg m_{b}$) \cite{kt}: \be\label{nb1}
n_{B}=\frac{g_{b}T^3}{6\pi^2}[\pi^2(\frac{\mu_{b}}{T})+(\frac{\mu_{b}}{T})^3]~,
\ee
 where $g_b$ is the number of intrinsic degrees of freedom of the
baryon. The final ratio of the baryon number to entropy is
\bea\label{nb2}
\left.\frac{n_{B}}{s}\right|_{T_D}\simeq-\frac{15g_b}{4\pi^2g_{\ast
s}}\left.\frac{c \dot R}{R T}\right|_{T_D}~, \eea where the cosmic
entropy density is $s=\frac{2\pi^2}{45}g_{\ast s}T^3$ and $g_{\ast
s}$ counts the total degrees of freedom of the particles which
contribute to the entropy of universe. $T_D$ in (8) is the
freezing out temperature of baryon number violation.

$\dot R/R$ in (\ref{nb2}) can be obtained from the Einstein
equation. For a constant equation of state of the fluid which
dominates the universe, one has \be
 \frac{\dot R}{R}=-3H(1+w)~.
 \ee
Hence, in the radiation-dominant epoch, $w=1/3$, and
 we have:
 \bea\label{dotrr}
 \frac{\dot R}{R}=-4H=-6.64g_{\ast}^{1/2}\frac{T^2}{m_{pl}}~.\label{reassume1}
 \eea
In the equation above we have used $H=\sim 1.66 g_{\ast}^{1/2}T^2/
m_{pl}$, and $g_{\ast}$ counts the total degrees of freedom of
effective massless particles. For most of the history of the
universe all particle species had a common temperature, and
$g_{\ast}$ is almost the same as $g_{\ast s}$, so, in the
following, we will not distinguish between them.

Substituting (\ref{dotrr}) into (\ref{nb2}) we arrive at a final
expression of the baryon number asymmetry: \bea
\left.\frac{n_{B}}{s}\right|_{T_D}&=&\frac{15}{\pi^2}\frac{c g_{b}H(T_D)}{g_{\ast}T_D}\nonumber\\
&\simeq &2.52 c g_{b} g_{\ast}^{-1/2} \frac{T_D}{m_{pl}}\sim
0.1c\frac{T_D}{m_{pl}}~. \eea In the numerical calculations above,
we have used $g_b\sim {\cal O}(1)$ and $g_{\ast}\sim {\cal
O}(100)$. Taking $c\sim {\cal O}(1)$, $n_B / s\sim 10^{-10}$
requires the decoupling temperature $T_D$ to be in the order of:
\be\label{result} T_D\sim 10^{-9} m_{pl}\sim 10^{10}~ \rm{GeV}~.
\ee A value of $T_D$ at or larger than $10^{10}$ GeV can be
achieved in theories of grand unification easily, however, if the
B-violating interactions conserve $B-L$, the asymmetry generated
will be erased by the electroweak Sphaleron \cite{manton}. In this
case $T_D$ will be as low as around 100 GeV and $n_B / s$
generated will be of the order of $10^{-18}$. Hence, now we turn
to leptogenesis \cite{lepto,zhang}. We take $ J^{\mu}$ in Eq.
(\ref{intlagr}) to be $J_{B-L}^\mu$. Doing the calculations with
the same procedure as above for $J^{\mu} = J^{\mu}_{B}$ we have
the final asymmetry of the baryon number minus lepton number
\begin{eqnarray}\label{fnumber2}
 \left.{n_{B-L}\over s}\right|_{T_D}\sim 0.1  c \frac{T_D}{m_{pl} }.
\end{eqnarray}
The asymmetry $n_{B-L}$ in (\ref{fnumber2})  will be converted to baryon
number
asymmetry when electroweak
Sphaleron $B+L$
interaction is in thermal equilibrium which happens for
temperature in the range of $10^2 ~{\rm GeV}
\sim 10^{12}{\rm GeV}$.  $T_D$ in (\ref{fnumber2}) is the temperature
below
which the $B-L$ interactions freeze out.

In the Standard Model of the electroweak theory, $B-L$ symmetry is
exactly conserved, however many models beyond the standard model,
such as Left-Right symmetric model predict the violation of the
$B-L$ symmetry. In this paper we take an effective Lagrangian
approach and parameterize the $B-L$ violation by higher
dimensional operators. There are many operators which violate
$B-L$ symmetry, however at dimension 5 there is only one
operator\footnote{Introducing a interaction between the dimension
5 operator in (14) and the Ricci scalar will induce the variation
of the neutrino masses during the evolution of the universe,
however the effect is negligible in this case \cite{gu,kaplan}.},
\begin{eqnarray}\label{lepvio}
{\cal L}_{\not L} = \frac{2} { f } l_L l_L \chi \chi +{\rm H.c.}~,
\end{eqnarray}
where $f$ is a scale of new physics beyond the Standard Model
which generates the $B-L$ violations, $l_L$ and $\chi$ are the
left-handed lepton and Higgs doublets respectively. When the Higgs
field gets a vacuum expectation value $< \chi > \sim v $, the
left-handed neutrino receives a majorana mass $m_\nu \sim
\frac{v^2}{f}$ .

In the early universe the lepton number violating rate induced by the
interaction in (\ref{lepvio}) is \cite{sarkar}
\begin{eqnarray}
  \Gamma_{\not L} \sim
    0.04~ \frac{T^3}{ f^2 }~.
\end{eqnarray}
Since $\Gamma_{\not L}$ is proportional to $T^3$, for a given $f$,
namely the neutrino mass, $B-L$ violation will be more efficient
at high temperature than at low temperature. Requiring this rate
be larger than the Universe expansion rate $\sim 1.66
g_{\ast}^{1/2}T^2/ m_{pl}$ until the temperature $T_D$, we obtain
a $T_D$-dependent lower limit on the neutrino mass:
\begin{eqnarray}
   \sum_i m_i^2  = {( 0.2 ~{\rm eV} ( { \frac{10^{12}~{\rm GeV}}{T_D} })^{1/2})}^2.
\end{eqnarray}
Taking three neutrino masses to be approximately degenerated,
~{\rm i.e.}~, $m_1 \sim m_2 \sim m_3\sim {\bar m}$ and defining
$\Sigma = 3 {\bar m}$, one can see that for $T_D \sim 10^{10}$
GeV, three neutrinos are expected to have masses $\bar{m}$ around
${\cal O}(1~{\rm eV})$. The current cosmological limit comes from
WMAP \cite{neutrino1} and SDSS \cite{sdss}. The analysis of Ref.
\cite{neutrino1} gives $\Sigma<0.69$ eV. The analysis from SDSS
shows, however that $\Sigma<1.7$ eV \cite{sdss}. These limits on
the neutrino masses requires $T_D$ be larger than $2.5\times
10^{11}$ GeV or $4.2\times 10^{10}$ GeV. The almost degenerate
neutrino masses required by the leptogenesis of this model will
induce a rate of the neutrinoless double beta decays accessible
for the experimental sensitivity in the near future \cite{beta}.
Interestingly, a recent study \cite{allen} on the cosmological
data showed a preference for neutrinos with degenerate masses in
this range.

The experimental CPT test with a spin-polarized torsion pendulum
\cite{cpt} puts strong limits on the axial vector background
$b_\mu$ defined by ${\cal L}=b_{\mu}{\bar e} \gamma^\mu \gamma_5
e$ \cite{colladay}: \be |{\vec b}| \leq 10^{-28} ~{\rm GeV}~. \ee
For the time component $b_0$, the bound is relaxed to be at the
level of $10^{-25}$ GeV \cite{mp}. In our model, assuming the
Ricci scalar couples to the electron axial current the same as Eq.
(\ref{intlagr}), we can estimate the CPT-violation effect on the
laboratory experiments. For a spatial-flat universe with a
constant equation of state of the dark energy $w_X$, one has \be
 \frac{\dot R}{R}=-3
H[1+\frac{w_X(1-3w_X)\Omega_X}{1-3w_X\Omega_X}]\sim -H~, \ee
 thus the current value of $\frac{\dot R_0}{R_0}$ is about $\sim
-H_0$, and the induced CPT-violating $b_0$ is \be b_0\sim
-c\frac{\dot R_0}{R_0}\sim  H_0\leq 10^{-42}~\rm{GeV}~, \ee which
is much below the current experimental limits.

In summary we have proposed a new type of interactions between the
Ricci scalar and the ordinary matter, and studied the possibility
of gravitational baryo(lepto)genesis. Our model can naturally
explain the baryon number asymmetry $n_B/s \sim 10^{-10}$ without
conflicting with the experimental tests on CPT.

{\bf{Acknowledgments:}} We would like to thank C.-G. Huang and B.
Wang for useful discussions. This work is supported in part by
National Natural Science Foundation of China under Grant Nos.
90303004  and 19925523 and by Ministry of Science and Technology
of China under Grant No. NKBRSF G19990754.

{}

\end{document}